\documentclass[
preprint,
prd,
aps,
showpacs,
tightenlines,
amsmath,
amssymb,
draft,
]{revtex4} 

\usepackage{dcolumn}
\usepackage{bm}
\input mathrsfs.sty

\def\D{{\mathscr D}}
\def\half{\frac{1}{2}}

\def\slash#1{\, /\kern-0.6em{#1}}

\begin{document}


\title{Monopole confinement by flux tube}     
\author{Chandrasekhar Chatterjee and Amitabha Lahiri}
\email{chandra@bose.res.in, amitabha@boson.bose.res.in}
\affiliation{S. N. Bose National Centre for Basic Sciences, \\
Block JD, Sector III, Salt Lake, Calcutta 700 098, INDIA\\ 
}

\begin{abstract}
We revisit Nambu's model of quark confinement by a tube of magnetic
flux, with two additional features. The quarks are taken to be
magnetic monopoles external to the tube, which seal the ends, and
are also taken to be fermions. This ensures that the model is
inconsistent unless there are at least two species of fermions
being confined.
\end{abstract}



\maketitle

Many years ago, Nambu suggested a model for the confinement of
quarks~\cite{Nambu:1974zg}, in which quarks are the endpoints of an
open Dirac string~\cite{Dirac:1948um}. If the gauge field is
massive, this string becomes real~\cite{Balachandran:1975qc},
unlike the usual Dirac string, which is a gauge artifact. In this
paper we make a small modification to Nambu's model -- instead of
treating the endpoints as quarks, we consider quarks as external
fermionic monopoles, which are then used to {\em seal} the
endpoints of the open string. This ensures that flux does not
escape or enter the endpoints, and a delta function(al) ensures
that quark currents always stick to the ends of strings. The string
itself is for us an infinitesimally thin tube of magnetic flux,
created in an Abelian Higgs model.

In the presence of strings, we first dualize the scalar field to
find the strings interacting via an antisymmetric tensor
potential~\cite{Kalb:1974yc, Davis:1988rw}, while the Abelian gauge
field is dualized to a `magnetic' photon~\cite{Mathur:1991ip,
Lee:1993ty}. Next we introduce fermionic magnetic monopoles into
the theory and minimally couple these to the magnetic photon. The
monopole current has to be axial, which produces an anomaly. We
cancel the anomaly by postulating additional species of fermionic
monopoles. Then we dualize the resulting theory again, to find a
theory of magnetic flux tubes interacting with a massive Abelian
vector gauge field. The tubes are sealed at the ends by fermions,
thus providing a toy model for quark confinement.

We start with the generating functional for the Abelian Higgs model
in $3+1$ dimensions, coupled to an Abelian gauge field $A^e_\mu$.
The partition function is given by
\begin{eqnarray}
Z &=& \int \D A^e_\mu\D\Phi\D\Phi^* \exp iS\,,\nonumber \\ S
&=& \int d^4 x \left(- \frac14 F^e_{\mu\nu}F^{e\mu\nu} +
|D_\mu\Phi|^2 + \lambda (|\Phi|^2 - v^2)^2 \right)\,,
\label{flux.Higgs}
\end{eqnarray}
where $D_\mu = \partial_\mu +i e A^e_\mu\,,$ and $F^e_{\mu\nu}
= \partial_\mu A^e_\nu - \partial_\nu A^e_\mu$ is the Maxwell
field strength.
    
In order to consider the situation where magnetic flux tubes are
present, we change variables from $\Phi\,,$$\Phi^*\,$ to the radial
Higgs field $\rho$ and the angular field $\theta\,,$ defined by  
$\Phi = \frac 1{\sqrt 2}\rho\exp(i\theta)\,.$ Then the measure
becomes, in these variables, 
\begin{eqnarray}
\int \D\Phi\D\Phi^*\cdots = \int \D\rho^2 \D\theta\cdots\,.
\label{flux.measure}
\end{eqnarray}
where the dots represent the measure for any other fields and the
integrand. Here we will consider the theory in the London limit
($\lambda\rightarrow \infty$ with the radial part of the field
fixed), i.e. we will ignore the $\rho$-dependent part of the
measure, and set $\rho=v$ (constant) in the action.

In the presence of flux tubes we can decompose the angular field
$\theta$ into a regular and a singular part\cite{Akhmedov:1995mw},
$\theta = \theta^r + \theta^s\,,$ where $\theta^s$ corresponds to a
given magnetic flux tube, and $\theta^r$ describes single valued
fluctuations around this configuration. The singular part of the
phase of the Higgs field is related to the world sheet $\Sigma$ of
the magnetic Abrikosov-Nielsen-Olesen string according to the
equation,
\begin{eqnarray}
\epsilon^{\mu\nu\rho\lambda}\partial_{\rho}
\partial_{\lambda}\theta^s &=& \Sigma^{\mu\nu}\,,  \\ 
\Sigma^{\mu\nu} &=&
2\pi n\int_{\Sigma}d\sigma^{\mu\nu}(x(\xi))\,\delta^4(x-x(\xi))\,,   
\label{flux.sigma}
\end{eqnarray}
where $\xi = (\xi^1, \xi^2)$ are the coordinates on the world-sheet
of the flux-tube, and $d\sigma^{\mu\nu}(x(\xi)) =
\epsilon^{ab}\partial_a x^\mu \partial_b x^\nu\,.$ In the above
equation $2\pi$ is the vorticity quantum in the units we are using
and $n$ is the winding number~\cite{Marino:2006mk}. The partition
function now has the form
\begin{equation}
Z = \int \D A^{e}_{\mu}\D\theta^s \D\theta^r \exp\left[i\int d^4 x
\left(- \frac14 F^e_{\mu\nu}F^{e\mu\nu} + \frac{v^2}2
(\partial_\mu\theta + e A^e_{\mu})^2 \right)\right]\,.
\label{flux.redZ}
\end{equation}

We can linearize the term $\displaystyle\frac{v^2}{2}
(\partial_{\mu}\theta + e A^e_{\mu})^2$ by introducing an auxiliary
field $C_\mu$ to get
\begin{eqnarray}
&& \int \D\theta^r \exp[i\int d^4 x \frac{v^2}{2}
(\partial_\mu\theta^r + \partial_\mu\theta^s + e A^e_{\mu})^2]\,
\nonumber \\ &&\qquad = \int \D\theta^r \D C_\mu \exp\left[-i\int d^4 x
\left\{ \frac 1{2v^2}\,C^2_\mu + C^\mu (\partial_\mu \theta^r + 
\partial_\mu \theta^s + e A^e_{\mu})\right\}\right]\, \nonumber \\
&&\qquad = \int \D C_\mu \, \delta[\partial_\mu C^\mu]
\exp\left[-i\int d^4x \left\{ \frac 1{2v^2}\,C^2_\mu + C^\mu
(\partial_\mu\theta^s + e A^e_\mu)\right\}\right]\,. 
\label{flux.Cmu}
\end{eqnarray}
We can resolve the constraint $\partial_\mu C^\mu = 0$ by
introducing an antisymmetric tensor field $B_{\mu\nu}$ and writing
$C_\mu$ in the form $C^\mu = \half
\epsilon^{\mu\nu\rho\lambda}\partial_\nu B_{\rho\lambda}$.
Integrating over the field $C_\mu\,,$ we get
\begin{eqnarray}
Z = \int \D A^e_{\mu} \D x_\mu(\xi) \D B_{\mu\nu} && \exp\left[i\int
d^4 x\left\{ -\frac14 F^e_{\mu\nu}F^{e\mu\nu}  \right.\right.\,
\nonumber \\ 
&&  \left.\left. + \frac{1}{12 v^2} H_{\mu\nu\rho}
H^{\mu\nu\rho} - \half \Sigma_{\mu\nu}B^{\mu\nu} - \frac e 2 
\epsilon^{\mu\nu\rho\lambda} A^e_\mu
\partial_{\nu}B_{\rho\lambda}\right\}\right]\,.
\label{flux.Cmuout}
\end{eqnarray}
Here we have written $H_{\mu\nu\rho} = \partial_\mu B_{\nu\rho} +
\partial_\nu B_{\rho\mu} + \partial_\rho B_{\mu\nu}$. We have also
replaced the integration over $\D\theta^s$ by an integration over
$\D x_\mu(\xi)$ which represents a sum over all configurations of
the worldsheet of the flux tube. Here $x_{\mu}(\xi)$ parametrizes
the surface on which the field $\theta$ is singular. The Jacobian
for this change of variables gives the action for the string on the
background spacetime~\cite{Akhmedov:1995mw}. The string has a
dynamics given by the Nambu-Goto action, plus higher order
operators~\cite{Polchinski:1991ax}, which can be obtained from the
Jacobian. We will not write the Jacobian explicitly in what
follows, but of course it is necessary to include it if we want to
study the dynamics of the flux tube.

Let us now integrate over the field $A^e_{\mu}$. To do this we have
to linearize $F^e_{\mu\nu}F^{e\mu\nu}$ by introducing another
auxiliary field $\chi_{\mu\nu}\,,$
\begin{eqnarray}
&& \int \D A^e_{\mu} \exp\left[i\int d^4x \left\{- \frac14
F^e_{\mu\nu}F^{e\mu\nu}-   
\frac e2 \epsilon^{\mu\nu\rho\lambda}A^e_\mu
\partial_{\nu}B_{\rho\lambda}\right\}\right]\,  \nonumber\\ 
&&\qquad = \int \D A^e_{\mu} \D \chi_{\mu\nu} \exp\left[i\int d^4x 
\left\{- \frac14 \chi_{\mu\nu}\chi^{\mu\nu} + \half
\epsilon^{\mu\nu\rho\lambda} \chi_{\mu\nu}
\partial_\rho A^e_\lambda - \frac e2 \epsilon^{\mu\nu\rho\lambda}
B_{\mu\nu}\partial_\rho A^e_\lambda \right\}\right]\, 
\nonumber \\  
&&\qquad = \int \D
\chi_{\mu\nu}\,\delta\Big[\epsilon^{\mu\nu\rho\lambda} 
\partial_{\nu}(\chi_{\rho\lambda}-e B_{\rho\lambda})\Big]\, 
\exp[i\int d^4x\{ -\frac 14 \chi_{\mu\nu}\chi^{\mu\nu}\}]\,.
\label{flux.Alinear}
\end{eqnarray}
We can integrate over $\chi_{\mu\nu}$ by solving the
$\delta$-functional as 
\begin{eqnarray}
\chi_{\mu\nu} = e B_{\mu\nu} + \partial_\mu A_\nu^m - \partial_\nu
A_\mu^m\,. 
\label{flux.Adual}
\end{eqnarray}
Here $A^m_{\mu}$ is the `magnetic photon', and what we have
achieved is dualization of the vector potential $A^e_{\mu}$. The
result of the integration is then inserted into
Eq.~(\ref{flux.Cmuout}) to give
\begin{equation}
Z = \int \D A^m_{\mu} \D x_{\mu}(\xi) \D B_{\mu\nu}
\exp\left[i\int\left\{- 
\frac 14 (e B_{\mu\nu} + \partial_{[\mu}A^m_{\nu]})^2
+ \frac 1{12 v^2} H_{\mu\nu\rho}H^{\mu\nu\rho} - 
\half \Sigma_{\mu\nu}B^{\mu\nu} \right\}\right] \,. 
\label{flux.functional}
\end{equation}

The equation of motion for the field $B_{\mu\nu}$ can be calculated
from this to be
\begin{equation}
\partial_\lambda H^{\lambda\mu\nu} = -\frac{m^2}e\, G^{\mu\nu} -
m^2 \,\Sigma^{\mu\nu} \,,
\label{flux.Beom} 
\end{equation}
where $G_{\mu\nu}= e B_{\mu\nu} + \partial_{\mu}A^m_{\nu} -
\partial_{\nu}A^m_{\mu}\,,$ and $m = ev$. Absence of magnetic current
gives $\partial_\mu G^{\mu\nu} = 0$, so from Eq.~(\ref{flux.Beom})
we get
\begin{equation}
\partial_\nu \Sigma^{\mu\nu} = 0\,.
\label{flux.closed}
\end{equation}
In particular, this equation means that in the absence of magnetic
monopoles, the vorticity tensor current $\Sigma_{\mu\nu}$ is
conserved. In other words, due to the conservation of magnetic flux
all the flux tubes in this case are closed or infinite. The
calculations so far have been done previously by others. What we
will do now has not appeared in the literature before this.

We now attach magnetic monopoles at the ends of a finite flux
tube. We will take the monopoles to be massless fermions and
minimally couple the monopole current to the magnetic or dual
photon. The monopole current must be axial for compatibility with
Maxwell's equations. After coupling, we will dualize the theory a
second time to get back to vector gauge fields, now coupled to flux
tubes.

However, a theory containing axial fermionic currents is anomalous
and if we try to dualize the theory, the presence of the anomaly
gives inconsistent results. We can cancel the anomaly by
introducing another species of fermionic monopoles with axial
charge opposite to the previous one. Let us denote the two species
by $q$ and $q'\,,$ with monopole charges $+g$ and $-g\,,$
respectively. So the monopole current becomes 
\begin{eqnarray}
j^\mu_m &=& g\bar q\gamma_5 \gamma^\mu q -
g\bar q'\gamma_5 \gamma^\mu q'\,.
\label{mono.anomaly}
\end{eqnarray}

The partition function of  Eq.~(\ref{flux.functional}) is modified
to include the fermionic monopoles, minimally coupled to the
`magnetic photon' $A_\mu^m\,,$ so the Lagrangian reads
\begin{eqnarray}
{\mathscr L} &=& - \frac 14 (e B_{\mu\nu} +
\partial_{[\mu}A^m_{\nu]})^2 + \frac 1{12 v^2}
H_{\mu\nu\rho}H^{\mu\nu\rho} - \half \Sigma_{\mu\nu}B^{\mu\nu}
+ i \bar q \slash\partial q + i \bar q' \slash\partial q' - A^m_\mu 
j^\mu_m \,.
\label{mono.lag}
\end{eqnarray}
The conservation condition of Eq.~(\ref{flux.closed}) is now
modified to 
\begin{eqnarray}
\frac 1e \partial_\nu \Sigma^{\mu\nu} =  j^\mu_m\,.
\label{mono.coneq}
\end{eqnarray}

This equation is a consequence of gauge invariance. To see this, we
take a transformation
\begin{eqnarray}
B_{\mu\nu} &\rightarrow& B_{\mu\nu} + \partial_{\mu}\Lambda_{\nu} -
\partial_{\nu}\Lambda_{\mu} \,, \nonumber \\ 
A^m_{\mu} &\rightarrow& A^m_{\mu} - \frac kg \Lambda_{\mu} \,.
\label{mono.gauge}
\end{eqnarray}
The second term of the Lagrangian of Eq.~(\ref{mono.lag}) is
invariant under the above transformation, while the first term is
made invariant by setting $eg = k\,.$ This is related to the Dirac
quantization condition as we shall see shortly. 

Since the flux due to the monopoles is fully confined in the tube,
the flux in the tube is $4\pi g$. Using
Eq.~(\ref{flux.sigma}), Eq.~(\ref{mono.coneq}) and the condition
$eg=k\,,$ we see that the "vorticity flux" must be
equal to flux of the monopoles. So we can write
\begin{eqnarray}
\label{flux}
2 n \pi = \mbox{vorticity flux}\;  = \frac kg (\mbox{monopole
flux})\; = 4\pi k
\end{eqnarray}

So $\displaystyle k = \frac n2$, and we must have the Dirac
quantization condition $\displaystyle eg = \frac n2\,,$ from our
assumption that the monopole flux is fully confined in the
tube. We can now write the partition function as
\begin{eqnarray}
Z[\Lambda_\mu] &=& \int \D A^m_{\mu} \D x_{\mu}(\xi) \D B_{\mu\nu}
\D q \D\bar q \D q' \D\bar q'\, \exp i\int d^4x \left[- \frac 14
(eB_{\mu\nu} + \partial_{[\mu}A^m_{\nu]})^2 \right. \nonumber\\ &&
\left. + \frac 1{12 v^2} H_{\mu\nu\rho}H^{\mu\nu\rho} - \half
\Sigma_{\mu\nu}B^{\mu\nu} -
\Sigma^{\mu\nu}\partial_{\mu}\Lambda_{\nu} +
e\Lambda_{\mu}j^{\mu}_m + i \bar q \slash\partial q + i \bar q'
\slash\partial q' - A^m_\mu j^\mu_m \right]\,.\; \nonumber \\
\label{mono.transpf}
\end{eqnarray}
Since (\ref{mono.gauge}) is only a change of variables, $Z$ cannot
depend on $\Lambda_\mu$. Thus $\Lambda_{\mu}$ can be integrated out
with no effect other than the introduction of an irrelevant
constant factor in $Z$, which we ignore. After integrating over
$\Lambda_\mu\,,$ we get
\begin{eqnarray}
Z = \int \D A^m_\mu\cdots
\delta\Big[\frac 1e \partial_\mu\Sigma^{\mu\nu} +  j^\nu_m\Big] 
&& \exp i\int d^4x \left[- \frac 14 (e B_{\mu\nu} +
\partial_{[\mu}A^m_{\nu]})^2  + \frac 1{12 v^2} H_{\mu\nu\rho}
H^{\mu\nu\rho} \right. 
\nonumber\\
&& \qquad\qquad \left. - \half\Sigma_{\mu\nu}B^{\mu\nu} 
+ i \bar q \slash\partial q + i \bar q'
\slash\partial q' - A^m_\mu j^\mu_m 
\right]\,, 
\end{eqnarray}
where the dots represent the measures for the other fields and
$x^\mu$.  One can see from the $\delta$-functional that the
vorticity current tensor is not conserved, but is cancelled by the
current of the added fermions. So the strings are open strings with
fermions stuck at the ends. Now we dualize the theory a second time
and get back to a vector gauge field.

Introducing an auxiliary field $\chi_{\mu\nu}$ to linearize the
first term of the Lagrangian, we get
\begin{eqnarray}
Z &=& \int \D A^m_\mu \cdots  \D \chi_{\mu\nu}\,
\delta\Big[\frac 1e\partial_\mu\Sigma^{\mu\nu} +  
j^\nu_m\Big] \exp\,i\int d^4x \Big[- \frac 14
\chi_{\mu\nu}\chi^{\mu\nu} + \half
\epsilon^{\mu\nu\rho\lambda}\chi_{\mu\nu} 
\partial_\rho A^m_\lambda\, \nonumber\\ 
&& \qquad + \frac 14 \epsilon^{\mu\nu\rho\lambda}
\chi_{\mu\nu}B_{\rho\lambda} +  
\frac 1{12 v^2} H_{\mu\nu\rho}H^{\mu\nu\rho}
- \half \Sigma_{\mu\nu}B^{\mu\nu} 
+ i \bar q \slash\partial q + i \bar q'
\slash\partial q' - A^m_\mu j^\mu_m 
\Big ] \,.
\end{eqnarray}
We can now integrate out $A^\mu_m$, and the result is
\begin{eqnarray}
Z = \int \D \chi_{\mu\nu} &\cdots
&\delta\Big[\frac 1e\partial_\mu\Sigma^{\mu\nu} +  j^\nu_m\Big]\,
\delta\Big[\half \epsilon^{\mu\nu\rho\lambda}
\partial_\nu\chi_{\rho\lambda} - j^\mu_m\Big]\,
\exp\, i\int d^4x \Big[- \frac14 \chi_{\mu\nu}\chi^{\mu\nu}
\nonumber\\   
&& + 
\frac e4 \epsilon^{\mu\nu\rho\lambda}\chi_{\mu\nu}B_{\rho\lambda} + 
\frac 1{12 v^2} H_{\mu\nu\rho}H^{\mu\nu\rho}
- \half \Sigma_{\mu\nu}B^{\mu\nu}  
+ i \bar q \slash\partial q + i \bar q'\slash\partial q' 
\Big ] \,.
\end{eqnarray}
Both the $\delta$-functionals must be satisfied, which requires
$\displaystyle \frac 1e \partial_\nu\Sigma^{\mu\nu} - \half
\epsilon^{\mu\nu\rho\lambda} \partial_\nu \chi_{\rho\lambda} =
0\,. $ This can be solved by introducing a gauge field $A_\mu$,
which allows the integration over $\chi_{\mu\nu}\,.$ Then the
partition function becomes
\begin{eqnarray}
Z &=& \int \D x_\mu(\xi) \D B_{\mu\nu} \D A_\mu\cdots \delta
\Big[\frac 1e\partial_\mu\Sigma^{\mu\nu} +  j^\nu_m\Big] \exp \,i\int d^4x
\Big[- \frac 14 F_{\mu\nu}F^{\mu\nu}
\nonumber\\ 
&& + \qquad \frac 1{12 v^2} H_{\mu\nu\rho} H^{\mu\nu\rho} + \frac
1{2g} \epsilon^{\mu\nu\rho\lambda} B_{\mu\nu} \partial_\rho
A_\lambda  
+ i \bar q \slash\partial q + i \bar q'\slash\partial q' \Big]\,.
\label{mono.final}
\end{eqnarray}
Here $F_{\mu\nu} = \partial_\mu A_\nu - \partial_\nu A_\mu -
\displaystyle\frac {1}{2e} \epsilon_{\mu\nu\sigma\lambda}
\Sigma^{\sigma\lambda}\,,$ the dots represent the fermion measure
and we continue to suppress the action for the flux tube itself, as
in Eq.~(\ref{flux.Cmuout}).

The theory is now in the form we originally intended, and contains
thin tubes of flux. The new feature is that the ends of the flux
tube are {\em sealed} by fermions, so that no flux escapes, all
flux is confined.  We should not think of this as any more than a
toy model of confinement, because the underlying theory is only the
Abelian Higgs model and not quantum chromodynamics. Even then, some
features are interesting enough to be highlighted.

There is a simple argument to calculate the length of the
string. The flux confined inside the string is $ 4 \pi g$, a
constant. The radius of the string core is of the order of
$1/v\sqrt{\lambda}\,.$ {}From this we can calculate the energy per
unit length of the tube to be $\mu \sim g^2 v^2 \lambda\,,$ also a
constant. Such a string, of finite length, would collapse in order
to minimize the energy unless it was stabilized by its angular
momentum.  For a rotating string of length $l\,,$ energy per unit
length $\mu,$ angular momentum $J\,,$ the energy function is $E =
\mu l + J^2/2\mu l^3\,.$ This has a minimum for the length
$L\sim\sqrt{J/\mu}\,.$ We see that for the stable flux tube with
magnetic monopoles at the ends,
\begin{equation}
\frac{J}{E^2} = constant\,,
\end{equation}
similar to the well-known Regge trajectory for mesons.

The gauge field $A_\mu$ is massive, with mass
$m=v/g$~\cite{Cremmer:1974mg, Allen:1990gb}. It does not couple
directly to the fermionic monopoles at the ends. Those fermions are
coupled only through the $\delta$-functional in
Eq.~(\ref{mono.final}), which guarantees that the monopoles must
seal the ends of the string. However, any other gauge field,
Abelian or not, axial or not, may be coupled to these fermions with
charge assignments independent of their charges under $A_\mu^m\,,$
which has been integrated out of the theory. In particular, if we
suggestively rename $q$ and $q'$ to $u$ and $\bar d\,,$ the allowed
configurations are $u\bar d\,, \bar u d\,,$ and $u\bar u \pm d\bar
d\,,$ which can couple to electroweak gauge fields. Note also that
we could have introduced three species of fermions (with charges
$1, 1, -2,$ for example) in Eq.~(\ref{mono.anomaly}), for the
purpose of anomaly cancellation, and we would again get flux tubes
with ends sealed by fermions. But a single species of fermions
would not produce such configurations.

In order to make the model more realistic, one would need to check
if flux is truly confined in the tube or if it escapes when the
tube has a finite thickness. A similar picture starting with an
axial gauge field and ending with a tube of `electric' rather than
magnetic flux will be interesting as well. Further, the freedom to
have other global symmetries in the theory allows in principle that
the $U(1)$ producing the string here may be embedded in an
$SU(N)_{global}\times U(1)_{local}$ symmetry, as
in~\cite{Vachaspati:1991dz}. We have not investigated these
questions, and also not the million-dollar question, which is
whether quantum chromodynamics has a low-energy effective sector
that behaves like an Abelian Higgs model.

\acknowledgments{}
It is a pleasure to acknowledege useful discussions with M.~Mathur,
P.~B.~Pal and R.~Banerjee.




\begin{thebibliography}{0}

\bibitem{Nambu:1974zg}
  Y.~Nambu,
  Phys.\ Rev.\ D {\bf 10}, 4262 (1974).


\bibitem{Dirac:1948um}
  P.~A.~M.~Dirac,
  Phys.\ Rev.\  {\bf 74}, 817 (1948).


\bibitem{Balachandran:1975qc}
  A.~P.~Balachandran, R.~Ramachandran, J.~Schechter, K.~C.~Wali and
H.~Rupertsberger, 
  Phys.\ Rev.\ D {\bf 13}, 354 (1976);
  Phys.\ Rev.\ D {\bf 13}, 361 (1976).


\bibitem{Kalb:1974yc}
  M.~Kalb and P.~Ramond,
  Phys.\ Rev.\ D {\bf 9}, 2273 (1974).


\bibitem{Davis:1988rw}
  R.~L.~Davis and E.~P.~S.~Shellard,
  Phys.\ Lett.\ B {\bf 214}, 219 (1988).

\bibitem{Mathur:1991ip}
  M.~Mathur and H.~S.~Sharatchandra,
  Phys.\ Rev.\ Lett.\  {\bf 66}, 3097 (1991).

\bibitem{Lee:1993ty}
  K.~M.~Lee,
  Phys.\ Rev.\ D {\bf 48}, 2493 (1993)

\bibitem{Akhmedov:1995mw}
  E.~T.~Akhmedov, M.~N.~Chernodub, M.~I.~Polikarpov and M.~A.~Zubkov,
  Phys.\ Rev.\ D {\bf 53}, 2087 (1996)

\bibitem{Marino:2006mk}
 E.~C. ~Marino,
 J.\ Phys.\ A {\bf 39}, L277 (2006).


\bibitem{Polchinski:1991ax}
  J.~Polchinski and A.~Strominger,
  Phys.\ Rev.\ Lett.\  {\bf 67}, 1681 (1991).


\bibitem{Cremmer:1974mg}
E.~Cremmer and J.~Scherk,
Nucl.\ Phys.\ B {\bf 72}, 117 (1974).


\bibitem{Allen:1990gb}
  T.~J.~Allen, M.~J.~Bowick and A.~Lahiri,
  Mod.\ Phys.\ Lett.\ A {\bf 6}, 559 (1991).


\bibitem{Vachaspati:1991dz}
  T.~Vachaspati and A.~Achucarro,
  Phys.\ Rev.\ D {\bf 44}, 3067 (1991).





\end{thebibliography}
\end{document}